\documentclass{an}
\usepackage{graphicx}
\usepackage{times}
\usepackage{fancyhdr}
\usepackage{amsmath}
\usepackage{amssymb}
\def\bmath#1{\boldsymbol{#1}}
\usepackage{natbib}

\def\secref#1{\hbox{\S\,\ref{#1}}}

\def\exref#1{(\ref{#1})}

\def\eqref#1{Eq.~(\ref{#1})}
\def\eqsref#1{Eqs.~(\ref{#1})}
\def\eqsand#1#2{Eqs.~(\ref{#1}) and~(\ref{#2})}
\def\eqsdash#1#2{Eqs.~(\ref{#1}-\ref{#2})}

\def\Eqsand#1#2{Equations~(\ref{#1}) and~(\ref{#2})}
\def\figref#1{Fig.~\ref{#1}}

\def\bea{\begin{eqnarray}}
\def\eea{\end{eqnarray}}

\def\({\left(}
\def\){\right)}
\def\<{\langle}
\def\>{\rangle}
\def\lt{\left}
\def\rt{\right}

\def\diff{\mathrm{d}}

\def\vdel{\bmath{\nabla}}

\def\vb{\bmath{\hat b}}

\def\vperp{v_\perp}

\def\vthi{v_{\mathrm{th}i}}

\def\vu{\bmath{u}}

\def\vB{\bmath{B}}

\def\dB{\delta B}

\def\mfp{\lambda_\mathrm{mfp}}
\def\mfpeff{\lambda_\mathrm{mfp,eff}}
\def\nueff{\nu_\mathrm{eff}}
\def\nusc{\nu_\mathrm{scatter}}
\def\nui{\nu_{ii}}
\def\gmax{\gamma_\mathrm{max}}

\def\Rm{\mathrm{Rm}}
\def\Re{\mathrm{Re}}
\def\Reeff{\Re_\mathrm{eff}}

\def\Beq{B_\mathrm{eq}}
\def\rhoeq{\rho_{i,\mathrm{eq}}}
\def\Oeq{\Omega_{i,\mathrm{eq}}}
\def\eps{\epsilon}
\def\Do{\Delta_0}

\def\pperp{p_\perp}
\def\ppar{p_\parallel}

\begin{document}

\title{Fast growth of magnetic fields in galaxy clusters: 
a self-accelerating dynamo}

\author{Alexander A.\ Schekochihin\inst{1} 
\and Steven C.\ Cowley\inst{2,3}}
\institute{
DAMTP, University of Cambridge, Wilberforce Rd, Cambridge CB3 0WA, UK
\and
Department of Physics and Astronomy, UCLA, Los Angeles, CA 90095-1547, USA
\and
Department of Physics, Imperial College London, 
Blackett Laboratory, Prince Consort Rd, London SW7 2BW, UK
}

\date{\today; e-print {\tt astro-ph/0508535}}

\abstract{We propose a model of magnetic-field growth in galaxy clusters 
whereby the field is amplified by a factor of about $10^8$ over a cosmologically 
short time of $\sim10^8$~yr. Our model is based on the idea that 
the viscosity of the intracluster medium during the field-amplification 
epoch is determined not by particle collisions but by plasma 
microinstabilities: these give rise to small-scale fluctuations, which 
scatter particles, increasing their effective collision 
rate and, therefore, the effective Reynolds number. 
This gives rise to a bootstrap effect as the growth of the field triggers 
the instabilities which increase the Reynolds number which, in turn, 
accelerates the growth of the field. The growth is explosive and the result 
is that the observed field strength is reached over a fraction of 
the cluster lifetime independent of the exact strength of the seed field 
(which only needs to be above $\sim10^{-15}$~G to trigger the explosive growth). 
}

\correspondence{as629@damtp.cam.ac.uk}

\maketitle

\addtocounter{page}{1}

\section{Introduction}

Rapidly improving measurements of Faraday Rotation in galaxy clusters 
reveal that intracluster medium is permeated by tangled magnetic fields 
with strength of a few $\mu$G and typical coherence scales of $1-10$~kpc 
\citep{Kronberg,Clarke_Kronberg_Boehringer,Carilli_Taylor,Govoni_Feretti_review,Vogt_Ensslin1,Vogt_Ensslin2}. This means that the intracluster medium (ICM) contains 
huge amounts of magnetic energy and it is natural to ask whether 
there is a robust and universal mechanism that could be 
responsible for generating such fields and that would also 
be insensitive to the particular circumstances in 
individual clusters (specific age of the cluster, magnitude of 
the seed field, etc.). We would like to forego 
the discussion of whether cluster fields could be purely of primordial 
or external origin \citep[as suggested by several authors; see, e.g.,][]{Kronberg_etal_agns,Banerjee_Jedamzik} and instead show that they can be amplified very quickly 
and with no stringent constraints on the magnitude of the seed field 
by the small-scale dynamo effect due to the turbulent motions of 
the ICM. The idea that the fields are dynamo generated has been explored by 
many authors \citep{Jaffe,Roland,Ruzmaikin_Sokoloff_Shukurov,DeYoung,Goldshmidt_Rephaeli,Kulsrud_etal_proto,SanchezSalcedo_Brandenburg_Shukurov,Subramanian_Shukurov_Haugen,Ensslin_Vogt_cores}. 
The main issues are whether there is, indeed, turbulence 
in clusters and whether the rms rate of strain associated with this turbulence 
is sufficiently large to exponentially grow the field to the observed strength 
in the cluster lifetime of a few Gyr. 

There are many possible energy sources in clusters that almost certainly 
produce random fluidlike motions of the ICM: merger events, 
subcluster and galactic wakes, active galactic nuclei. Various 
indirect observational estimates \citep[e.g.,][]{Schuecker_etal,Rebusco_etal,Fujita} 
appear to converge in expecting random flows with velocity fluctuations 
$U \sim 10^2-10^3$~km/s at the outer scale $L\sim10^2-10^3$~kpc 
\citep[a direct detection may be achieved in the near future; see][]{Inogamov_Sunyaev}. 
Similar numbers are obtained in numerical simulations 
of cluster formation \citep{Norman_Bryan,Sunyaev_Norman_Bryan,Ricker_Sarazin,Takizawa}. 
There is, however, no consensus on whether turbulence, at least in the usual 
hydrodynamic sense of a Kolmogorov energy cascade across a broad inertial range of 
scales, is a generic feature of clusters \citep{Fabian_etal_Perseus2}. 
The main difficulty is the very large values of 
the ICM viscosity obtained via the standard estimate $\sim\vthi\mfp$, 
where $\vthi\sim10^3$~km/s is the ion thermal speed and $\mfp\sim1-10$~kpc 
is the mean free path (assuming fully ionised hydrogen ICM with 
temperature $T\sim10^8$~K and density $n\sim10^{-3}$~cm$^{-3}$). 
This gives $\Re\sim10^2$ if not less, which makes the existence of a 
well-developed inertial range doubtful. 

In fact, the small-scale dynamo does not 
require a turbulent velocity field in the sense of a broad 
inertial range. As the dynamo effect is simply the random stretching of field lines, 
it is always controlled by the rms rate of strain --- and this is associated with 
the viscous-scale motions, which are random but spatially smooth. 
Whether the viscous scale ($\sim L\Re^{-3/4}$) is much smaller than 
the outer scale $L$ is inessetial as long as random motions are 
maintained\footnote{See numerical simulations of small-scale dynamo 
with $\Re\sim1-10^2$ by \citet{SCTMM_stokes,Haugen_Brandenburg_Dobler} 
and the review of the astrophysical turbulence and dynamo by \citet{SC_mhdbook}.} 
--- not a problem in clusters, given the variety of available stirring 
mechanisms. Thus, the field will be amplified and the only issue is whether 
this happens sufficiently fast, given the relatively small growth rate 
obtainable in a viscous ICM. We shall see that this turns out not to be 
a problem because the effective ICM viscosity in the weak-field regime 
is much smaller than $\vthi\mfp$. 

The widely used viscosity estimate $\sim\vthi\mfp$ only applies 
when the magnetic field is so weak that 
that the ion gyroradius is larger than the mean free path: $\rho_i>\mfp$. 
This gives $B\lesssim10^{-18}$~G --- much smaller than the observed field 
in clusters. For stronger fields, the plasma becomes magnetised and its 
viscosity anisotropic: the standard value still works parallel to the magnetic 
field, while the motions perpendicular to the field are virtually 
undamped \citep{Braginskii}. As was pointed out recently by \citet{SCKHS_brag}, 
the anisotropy of the viscous stress in clusters means anisotropy of 
the plasma pressure, which, in turn, always gives rise to 
well-known plasma microinstabilities \citep[most importantly, 
mirror and firehose, see, e.g.,][]{Gary_book}
at scales all the way down to the ion gyroscale. The emergence of the 
pressure anisotropies is intimately related to the changes in the strength 
of the magnetic field: in a magnetised plasma, the conservation of 
the first adiabatic invariant for each particle, $\mu=\vperp^2/B$, 
implies that the perpendicular plasma pressue changes 
according~to\footnote{Here and in the rest of this paper, 
``$=$'' really means ``$\sim$''.} 
\bea
{1\over\pperp}{\diff\pperp\over\diff t} = 
{1\over B}{\diff B\over\diff t} - \nui\,{\pperp-\ppar\over\pperp},
\eea
where the first term is due to the conservation of $\mu$ and 
the second term represents collisional isotropisation 
($\nui\sim\vthi/\mfp$ is the ion collision frequency). 
Balancing the two terms, we get 
\bea
{1\over B}{\diff B\over\diff t} = \nui\Delta, 
\label{mu_cons}
\eea
where $\Delta = (\pperp-\ppar)/\pperp$. 
On the other hand, since the resistivity of the ICM is tiny 
($\Rm\sim10^{29}$), the field lines are frozen into the fluid flow 
and the change of the magnetic-field strength 
can be expressed in terms of the fluid velocity $\vu$:
\bea
{1\over B}{\diff B\over\diff t} = \vb\vb:\vdel\vu - \vdel\cdot\vu 
= {U\over L}\,\Re^{1/2},
\label{ind_eq}
\eea 
where $\vb=\vB/B$. 
The last equality in \eqref{ind_eq} is based on estimating the 
turbulent rate of strain at the viscous scale. It is the parallel 
viscosity that matters here because the motions that 
change $B$ are precisely the ones --- and the only ones --- that 
are damped by it \citep{Braginskii}. 
Since 
\bea
\label{Re_def}
\Re={UL\over\vthi\mfp}={UL\over\vthi^2}\,\nui, 
\eea
we have, from \eqsand{mu_cons}{ind_eq}, 
\bea
\label{Delta_def}
\Delta = {U\over\vthi}\({U\over L}{1\over\nui}\)^{1/2} 
= \(U\over\vthi\)^2\Re^{-1/2}.
\eea
The peak growth rate of these instabilities (at the ion gyroscale)~is
\bea
\gmax = \(|\Delta|-{2\over\beta}\)^{\alpha_1}\Omega_i,
\eea
where $\beta=8\pi nT/B^2$, $\Omega_i=eB/cm_i$ is the ion cyclotron 
frequency, $\alpha_1=1$ for the mirror and $\alpha_1=1/2$ for 
the firehose instability. To sum up briefly, external energy sources 
drive random motions, which change the field [\eqref{ind_eq}], 
which gives rise to pressure anisotropies 
[\eqref{mu_cons}], which trigger the instabilities. 
The latter are stabilised if the magnetic field is sufficiently 
strong: $\beta<2/\Delta$. 

It is not currently clear exactly how the instabilities affect the 
structure of the turbulence and magnetic fields in clusters. 
One plausible hypothesis is that the saturation is quasilinear 
with small fluctuations $\dB$ at the gyroscale: 
\bea
{\dB^2\over B^2} = \(|\Delta|-{2\over\beta}\)^{\alpha_2},
\eea
where $\alpha_2$ is some positive power. 
These fluctuations will scatter particles, giving an effective collision 
frequency: 
\bea
\nusc = {\dB^2\over B^2}\,\gmax 
= \(|\Delta| - {2\over\beta}\)^\alpha\Omega_i,
\label{nusc_def}
\eea 
where $\alpha = \alpha_1+\alpha_2>0$. 
The result is an effective Reynolds number [\eqref{Re_def} with $\nui\to\nusc$] 
that is much larger than the one based on the actual ion-ion collisions and that 
increases  with the magnetic-field strength 
(because $\nusc\propto\Omega_i\propto B$). 
Since this, in turn, speeds up the small-scale dynamo by increasing 
the turbulent rate of strain [\eqref{ind_eq}], we expect explosive 
growth of magnetic field. This possibility was briefly referred to in 
\citet{SCKHS_brag}. In \secref{sec_model}, we shall construct a simple 
model of such a self-accelerating dynamo 
and evaluate its effectiveness in generating observed cluster 
fields for a set of fiducial cluster parameters (introduced in \secref{sec_numbers}). 
We will conclude with a summary and some words of caution in~\secref{sec_conc}. 

\section{Cluster parameters}
\label{sec_numbers}

In order to makes specific estimates as we proceed, let us assume a fiducial 
set of parameters for the ICM:\footnote{We emphatically do not indend 
to suggest that the actual values 
of these quantities are known for real clusters to more than an 
order-of-magnitude precision (in some cases, even worse than that). 
These numbers are merely a convenient example.} 
\bea
&&n = 10^{-3}~\mathrm{cm}^{-3},\\
&&T = 10^8~\mathrm{K}\quad\Rightarrow\quad
\vthi = \sqrt{T\over m_i}\simeq 1000~\mathrm{km/s},\\
&&U = 300~\mathrm{km/s}\quad\mathrm{(rms~turbulent~velocity)},\\
&&L = 200~\mathrm{kpc}\qquad\!\mathrm{(outer~scale)},\\
&&\mfp = 1~\mathrm{kpc}.
\eea 
We shall measure magnetic field in units of 
\bea
\label{Beq_def}
\Beq = (8\pi n T)^{1/2} \simeq 20~\mu\mathrm{G},
\eea
for which $\beta(\Beq)=1$. For this field, the ion cyclotron 
frequency and the ion gyroradius are 
\bea
\Oeq &=& {e\Beq\over cm_i}\simeq 0.2~\mathrm{s}^{-1},\\ 
\rhoeq &=& {\vthi\over\Oeq}\simeq 5000~\mathrm{km}. 
\eea
The magnetic field strength above which plasma is magnetised 
($\rho_i<\mfp$)~is
\bea
\label{B0_def}
B_0 = \Beq\,{\rhoeq\over\mfp} \sim 10^{-18}~\mathrm{G}. 
\eea
The turbulence is characterised by $\Re\simeq60$ and 
\bea
\label{outert_def}
&&{L\over U} \sim 10^9~\mathrm{yr}\qquad\qquad\qquad\ \mathrm{(outer~timescale)},\\
\label{shear_def}
&&{L\over U}\,\Re^{-1/2} \sim 10^8~\mathrm{yr}\qquad\quad\mathrm{(viscous~timescale)},\\
&&l_{\nu}\sim L\,\Re^{-3/4}\sim10~\mathrm{kpc}\quad\,\mathrm{(viscous~cutoff)}. 
\eea

\section{A model of self-accelerating cluster dynamo}
\label{sec_model}

\begin{figure*}
{
\begin{tabular}{ccc}
(a) & & (b)\\
\includegraphics[height=5cm]{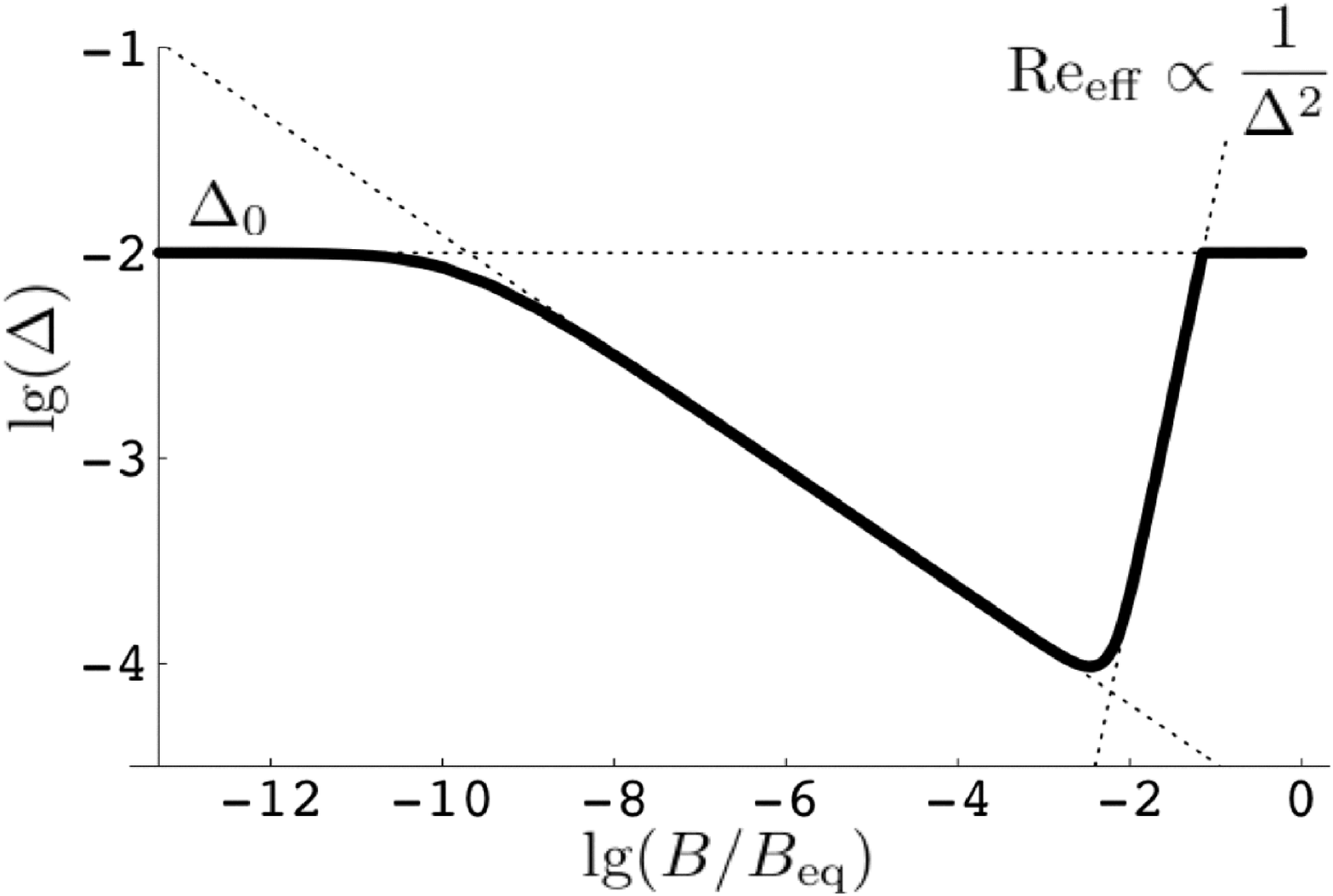}&\qquad&\includegraphics[height=5cm]{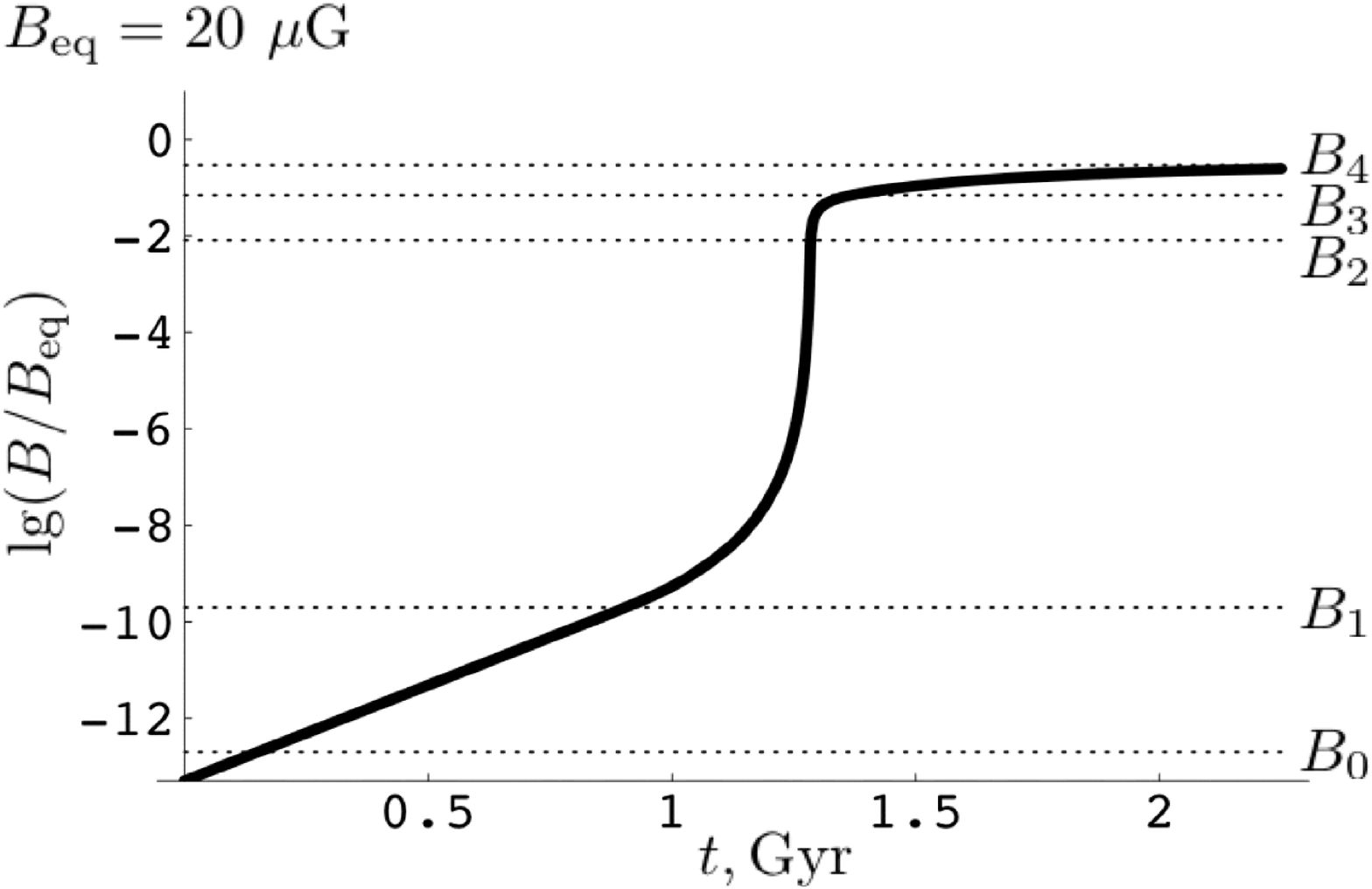}
\end{tabular}
}
\caption{(a) The anisotropy ratio vs.\ field strength. Dotted lines 
are the analytical solutions for the three asymptotic regimes 
discussed in \secref{sec_exponential}, \secref{sec_expl}, and \secref{sec_alg} 
[$\Delta=\Do$, $\Delta(B)$ given by \eqref{Delta_expl}, and $\Delta(B)=2B^2$]. 
(b) Growth of the magnetic field. 
The dotted lines show various field-strength thresholds 
discussed in the text [\eqsref{B0_def}, \exref{B1_def}, \exref{B2_def}, \exref{B3_def} 
and \exref{B4_def}]. The initial slope of the curve is the rate at which 
the conventional dynamo can grow the field. 
Both plots are for $\alpha=3/2$ and the fiducial set of 
cluster parameters given in \secref{sec_numbers}.} 
\label{fig_DandB}
\end{figure*}

We assume that the net effect of the instabilities is 
to change the effective collision frequency of ions:
\bea
\nueff = \nui + \nusc(B),
\eea
where $\nusc$ is given by \eqref{nusc_def}. 
If we use \eqsdash{Re_def}{Delta_def} with $\nui$ replaced 
by $\nueff$, we find the following equation for the anisotropy ratio 
\bea
{1\over\Delta^2} = {1\over\Do^2} + 
{1\over\eps}\(\Delta - 2B^2\)^\alpha B,
\label{Delta_eq}
\eea
where magnetic field is in units of $\Beq$ [\eqref{Beq_def}] and 
\bea
\Do &=& \lt[\({U\over\vthi}\)^3{\mfp\over L}\rt]^{1/2} \sim 0.01,\\ 
\eps &=& \({U\over\vthi}\)^3{\rhoeq\over L} = {\rhoeq\over\mfp}\,\Do^2 
\sim 10^{-17}.
\eea
Our results will not be very sensitive to the value of $\alpha$. 
In what follows, we shall use $\alpha=3/2$ for specific estimates. 

As the instabilities grow much faster than the field changes, 
we can assume that, as far as the field is concerned,  
the adjustament of the collision frequency occurs instantaneously. 
The growth of the magnetic field satisfies [from \eqsand{ind_eq}{Delta_def}
with $\nui\to\nueff$]
\bea
\label{B_eq}
{1\over B}{\diff B\over\diff t} = {1\over\Delta(B)},
\eea
where time is in units of $(\vthi/U)^2L/U$ 
and $\Delta(B)$ is the solution of \eqref{Delta_eq}. 
\Eqsand{Delta_eq}{B_eq} constitute our model of the field growth. 
They are solved anlytically in three asymptotic regimes 
in what follows and numerically in \figref{fig_DandB}. 

\subsection{Exponential stage}
\label{sec_exponential}

Let us start with an arbitrarily small seed field. As long as the 
the first term in \eqref{Delta_eq} dominates, the effect of 
the instabilities is negligible and the field grows exponentially 
with the characteristic $e$-folding time $\sim\Do$, which 
in dimensional units translates 
to the inverse rms rate of strain $\sim10^8$~yr [\eqref{shear_def}]. 
This initial dynamo stage continues until the second term 
in \eqref{Delta_eq} becomes comparable to the first, 
which occurs at the field strength 
\bea
\label{B1_def}
B_1 = {\eps\over\Do^{2+\alpha}} 
= \Beq\,{\rhoeq\over\mfp}\lt[\biggl({\vthi\over U}\biggr)^3
{L\over\mfp}\rt]^{\alpha/2}. 
\eea
For $\alpha=3/2$, we have $B_1\sim 10^{-15}~\mathrm{G}$. 

\subsection{Explosive stage} 
\label{sec_expl}

For $B\gg B_1$, we may approximate the solution of 
\eqref{Delta_eq}~by
\bea
\label{Delta_expl}
\Delta(B) \simeq \({\eps\over B}\)^{1/(2+\alpha)}.
\eea
Upon substitution into \eqref{B_eq}, this gives an explosively 
growing field:
\bea
B(t) = \epsilon\(2+\alpha\over t_c-t\)^{2+\alpha}, 
\eea
where, taking $B(0)=B_1$, we get $t_c=(2+\alpha)\Do$. 
During this very fast stage, the anisotropy drops precipitously, while 
the effective Reynolds number and, therefore, the rms rate of strain, 
grow:
\bea
\label{Reeff_exp}
\Reeff = \({U\over\vthi}\)^4{1\over\Delta(B)^2}.
\eea 
Hence the self-accelerating nature of the dynamo. 
Eventually (in fact, very soon), the field grows to be strong enough 
so that $B^2$ is comparble to $\Delta(B)$, the second term in 
\eqref{Delta_eq} starts dropping again and the $\Delta$ stops 
decreasing and starts increasing. 
The minimum value of $\Delta$ is reached at $B\sim B_2$, where 
\bea
\label{B2_def}
B_2 = \eps^{1/(5+2\alpha)}
= \Beq\lt[\({U\over\vthi}\)^3{\rhoeq\over L}\rt]^{1/(5+2\alpha)}.
\eea
For $\alpha=3/2$, we have $B_2\sim10^{-7}~\mathrm{G}$. 
The time it takes to achieve this is 
\bea
t\simeq t_c-(2+\alpha)\,\epsilon^{2/(5+2\alpha)} \simeq t_c 
\sim10^8~\mathrm{yr}.
\eea
Over this time, the field is amplified by a factor 
of $\sim10^8$. This is to be compared with the conventional dynamo 
\citep[e.g.,][]{Subramanian_Shukurov_Haugen}, which has the $e$-folding 
time of $10^8$~yr and would, therefore, need around $2$~Gyr to achieve 
an amplification factor of $10^8$. 

\subsection{Algebraic stage}
\label{sec_alg}

When $B>B_2$, $\Delta$ stays just above $2B^2$. Computing the small 
correction, we obtain the following solution of \eqref{Delta_eq}
\bea
\Delta(B) \simeq 2B^2 +\lt[{\eps\over B}
\({1\over4 B^4}-{1\over\Do^2}\)\rt]^{1/\alpha}.
\eea
Solving \eqref{B_eq} with $\Delta\simeq2B^2$ and $B(0)=B_2$, we get 
\bea
B(t) = \sqrt{B_2^2+t}.
\eea
This is a rather slow growth but it only lasts until the first term 
in \eqref{Delta_eq} is again dominant: the transition is, in fact, 
sharp because once $\Delta<2B^2$, the instabilities are shut down 
and we have to set $\nusc=0$ for all greater values of $B$. 
The field at which this happens is 
\bea
\label{B3_def}
B_3 = \({\Do\over2}\)^{1/2} = 
{\Beq\over\sqrt{2}}\lt[\({U\over\vthi}\)^3{\mfp\over L}\rt]^{1/4} 
\sim 1~\mu\mathrm{G}.
\eea
The time to get there is $t\simeq B_3^2=\Do/2\sim10^8$~yr again. 
The amplification factor since the end of the explosive stage is $\sim10$. 

Note that that if we compare $\rho_i$ with the effective mean free path
$\mfpeff=\vthi/\nueff$, we get 
\bea
{\rho_i(B)\over\mfpeff(B)} = {\Beq\over B}{\eps\over\Delta(B)^2}, 
\eea
which is readily seen to decrease throughout all of the dynamo stages discussed 
above. Therefore, the magnetised-plasma assumption, once true at $B=B_0$, 
is never broken. For example, at the point of the largest effective Reynolds 
number ($\Re\sim10^6$ at end of the explosive regime, $B=B_2$)
we have $\mfpeff\simeq0.03$~pc and $\rho_i\sim10^6$~km. 

\subsection{Nonlinear stage}

So far, the dynamo we have considered has been kinematic, i.e., 
the magnetic field has remained too weak to exert a back reaction 
on the flow. The condition for this to be true is that 
the field energy remains smaller than the energy of the smallest-scale 
turbulent motions that can change the field, i.e., the viscous-scale 
motions \citep[see][]{SCTMM_stokes}. 
In Kolmogorov turbulence, their velocity is $U\,\Re^{-1/4}$, 
whence, using \eqsand{Reeff_exp}{Beq_def}, we find that the 
condition for the nonlinearity to become important is 
$B^2 \gtrsim \Delta(B)$. 
Thus, all through the algebraic regime, the magnetic field hovers 
around this threshold. After instabilities are stabilised, it 
can finally exceed it by growing above $B_3$. For $B>B_3$, we have 
to model the effect of the back reaction. How to do this is a highly 
nontrivial problem for which a definitive solution has not so far been found. 
Here, we shall use the model proposed by 
\citet{SCHMM_ssim},\footnote{For numerical evidence in support 
of the model, see \citet{SCTMM_stokes}.}
which is constructed in a somewhat similar spirit as the 
self-accelerating dynamo model above. The main idea is that 
the growing magnetic field gradually suppresses the ability of 
turbulent eddies to stretch it. At any given point in time, all 
eddies whose energy is less than the energy of the field are thus 
suppressed. Final saturated state is reached when the energy of the 
field is comparable to the energy of the outer-scale eddies, i.e., 
to the total energy of the fluid turbulence: 
\bea
\label{B4_def}
B_4 = \Beq\,{U\over\vthi} \simeq 6~\mu\mathrm{G}.
\eea
To model this process, we assume that the magnetic field evolves 
according~to 
\bea
\label{B_eq_nlin}
{1\over B}{\diff B\over\diff t} = \gamma(B),
\eea
with the $B$-dependent growth rate \citep{SCHMM_ssim}
\bea
\nonumber
\gamma(B) &=& {U\over L}\,\Re^{1/2}\lt[1-{1\over\(1+B_4^2/B_3^2\)^2}\rt]^{-1/2}\\
&&\times\lt[{1\over\(1+B^2/B_3^2\)^2}-{1\over\(1+B_4^2/B_3^2\)^2}\rt]^{1/2}.
\label{gamma_def}
\eea
It is not hard to see that, for $B_3\ll B\ll B_4$, $\gamma(B)\propto1/B^2$, 
which again gives the magnetic field growing as $B\sim\sqrt{t}$. 
Thus, the slow algebraic growth continues as the field exceeds $B_3$
until it saturates at $B=B_4$. Again, it does not take long because 
the inertial range in clusters is very narrow and the field does have 
to grow very much to get from $B_3$ to $B_4$. 

\section{Conclusions}
\label{sec_conc}

All stages of the evolution of the magnetic field described in the previous 
section are illustrated in \figref{fig_DandB}, which presents the numerical 
solution of \eqsand{B_eq}{Delta_eq} [followed by \eqsdash{B_eq_nlin}{gamma_def} 
in the nonlinear stage]. We see that the field grows 
by a factor of about $10^{12}$ over a time period of $\sim2$~Gyr. 
Most of the growth occurs over a short fraction of this time 
during the explosive self-accelerating dynamo stage, which takes the field 
from any value above $\sim10^{-15}$~G to $\sim10^{-7}$~G in a 
cosmologically short time of about $10^8$~yr. After that, the field 
does not have to grow further for a very long time before it reaches 
the strength of several $\mu$G consistent with the observational data. 
The importance of the self-accelerating dynamo mechanism we have proposed 
is that it provides for field amplification 
to the observed strength over a short period of 
time {\em independent of the precise value of the seed field\footnote{For seed 
fields below $\sim10^{-15}$~G, the conventional exponentially growing 
dynamo with $e$-folding time $\sim10^8$~yr is required until 
$10^{-15}$~G is reached.}} 
or, indeed, of exactly how many Gyr a cluster has been around. 
This removes the need to envision ways of generating seed fields 
satisfying the often quite large lower bounds that 
theories (cited in the Introduction) 
and numerical simulations \citep[e.g.,][]{Roettinger_Stone_Burns,Dolag_Bartelmann_Lesch1,Dolag_Bartelmann_Lesch2,Dolag_etal_mag,Brueggen_etal_mag} 
infer to be necessary for 
magnetic fields of observed strength to be produced by the conventional 
turbulent dynamo over the cluster lifetime. 
Since the amplification time is so short, it is also 
probably not crucial whether the cluster turbulence responsible for 
amplification is best modelled as forced or decaying 
\citep[cf.][]{Subramanian_Shukurov_Haugen}. 
We conclude that, no matter what the seed field is, 
the random motions in the ICM will have no difficulty in amplifying 
it to the observed level in a fraction of the cluster lifetime. 

Finally, we must alert the reader to an important 
issue, which we have ignored above. This concerns the 
spatial structure of the cluster fields. Theory and simulations 
of the small-scale dynamo in the MHD description indicate that the 
dynamo-generated fields have a folded structure with 
direction reversals at the resistive scale $\sim L\,\Rm^{-1/2}$ 
\citep{Ott_review,SCTMM_stokes,Brandenburg_Subramanian}. 
In clusters, this scale turns out to be extremely small 
($\sim10^4$~km), while analysis of the observational data suggests 
that the typical reversal scale is $\sim1$~kpc \citep{Vogt_Ensslin1,Vogt_Ensslin2}. 
It is likely that the reversal scale is, in fact, set by some form of anomalous 
resistivity associated with electron scattering by the plasma 
instabilities discussed above. We relegate further development of 
this idea to future work. 

\acknowledgements
It is a pleasure to acknowledge helpful discussions with T.~En{\ss}lin, 
G.~Hammett, R.~Kulsrud, E.~Quataert, P.~Sharma, A.~Waelkens. 
A.A.S.\ was supported by a UKAFF Fellowship. 
This work has benefited from research funding from the EU Sixth 
Framework Programme under RadioNet R113CT 2003 5058187 
and from the US DOE Center for Multiscale Plasma Dynamics, 
Fusion Science Center Cooperative Agreement ER54785.

\end{document}